\documentclass[12pt,preprint]{aastex}
\usepackage{graphicx,amssymb,natbib,longtable,lscape}
\usepackage{cite}
\usepackage{amsmath}

\newcommand{\about}{$\sim$}
\newcommand{\kms}{km s$^{-1}$}

\newcommand{\etal}{et al.}

\newcommand{\degrees}{\ensuremath{^\circ}}

\newcommand{\Halpha}{H$\alpha$}
\newcommand{\HI}{\mbox{H\textsc{i}}}

\newcommand{\SigmaQ}{\ensuremath{\Sigma_Q}}
\newcommand{\SigmaA}{\ensuremath{\Sigma_A}}
\newcommand{\SigmaS}{\ensuremath{\Sigma_\text{S04}}}
\newcommand{\solarmassespersquareparsec}{M$_\odot$ pc$^{-2}$}

\newcommand{\Htwo}{H$_2$}
\newcommand{\per}[1]{\ensuremath{^{-#1}}}

\shortauthors{Hallenbeck \etal}

\begin{document}
\shorttitle{HIghMass Galaxies UGC 9037 and 12506}
\title{HIghMass - High HI Mass, HI-rich Galaxies at z$\sim0$\\High-Resolution VLA Imaging of UGC 9037 and UGC 12506}
\author{Gregory Hallenbeck\altaffilmark{1}, Shan Huang\altaffilmark{1,2}, Kristine Spekkens\altaffilmark{3}, Martha P. Haynes\altaffilmark{1}, Riccardo Giovanelli\altaffilmark{1}, Elizabeth A. K. Adams\altaffilmark{1,4}, Jarle Brinchmann\altaffilmark{5}, Jayaram Chengalur\altaffilmark{6}, Leslie K. Hunt\altaffilmark{7}, Karen L. Masters\altaffilmark{8,9}, Am\'{e}lie Saintonge\altaffilmark{10}}
\altaffiltext{1}{Center for Radiophysics and Space Research, Space Sciences Building, Cornell University, Ithaca, NY 14853. {\textit{e-mail:}} ghallenbeck@astro.cornell.edu, haynes@astro.cornell.edu, riccardo@astro.cornell.edu}
\altaffiltext{2}{Academia Sinica, Institute of Astronomy \& Astrophysics, P.O. Box 23-141, Taipei 10617, Taiwan. {\textit{e-mail:}} shan@asiaa.sinica.edu.tw}
\altaffiltext{3}{Royal Military College of Canada, Department of Physics, PO Box 17000, Station Forces, Kingston, Ontario, Canada K7K 7B4. {\textit{e-mail:}} Kristine.Spekkens@rmc.ca}
\altaffiltext{4}{ASTRON/Netherlands Institute for Radio Astronomy, Oude Hoogeveensedijk 4, 7991 PD Dwingeloo, The Netherlands. {\textit{e-mail:}} adams@astron.nl}
\altaffiltext{5}{Leiden Observatory, Leiden University, P.O. Box 9513, 2300 RA Leiden, Netherlands {\textit{e-mail:}} jarle@strw.leidenuniv.nl}
\altaffiltext{6}{National Centre for Radio Astrophysics, Tata Institute for Fundamental Research, Pune 411 007, India. {\textit{e-mail:}} chengalu@ncra.tifr.res.in}
\altaffiltext{7}{INAF-Osservatorio Astrofisico di Arcetri, Largo E. Fermi 5, I-50125, Firenze, Italy. {\textit{e-mail:}} hunt@arcetri.inaf.it}
\altaffiltext{8}{Institute of Cosmology and Gravitation, Dennis Sciama Building, Burnaby Road, Portsmouth POI 3FX. {\textit{e-mail:}} Karen.Masters@port.ac.uk}
\altaffiltext{9}{South East Physics Network, www.sepnet.ac.uk}
\altaffiltext{10}{University College London Dept. of Physics \& Astronomy, Kathleen Lonsdale Building, Gower Place, London, WC1E 6BT, United Kingdom. {\textit{e-mail:}} a.saintonge@ucl.ac.uk}

\begin{abstract}
\noindent
We present resolved \HI\ observations of two galaxies, UGC 9037 and UGC 12506, members of a rare subset of galaxies detected by the ALFALFA extragalactic \HI\ survey characterized by high \HI\ mass and high gas fraction for their stellar masses. Both of these galaxies have M$_*>10^{10}$ M$_\odot$ and M$_\text{HI}>$ M$_*$, as well as typical star formation rates for their stellar masses. How can such galaxies have avoided consuming their massive gas reservoirs? From gas kinematics, stability, star formation, and dark matter distributions of the two galaxies, we infer two radically different histories. UGC 9037 has high central \HI\ surface density ($>10$ \solarmassespersquareparsec). Its gas at most radii appears to be marginally unstable with non-circular flows across the disk. These properties are consistent with UGC 9037 having recently acquired its gas and that it will soon undergo major star formation. UGC 12506 has low surface densities of \HI, and its gas is stable over most of the disk. We predict its gas to be \HI-dominated at all except the smallest radii. We claim a very high dark matter halo spin parameter for UGC 12506 ($\lambda=0.15$), suggesting that its gas is older, and has never undergone a period of star formation significant enough to consume the bulk of its gas.

\end{abstract}

\section{Introduction}


\noindent
In the standard hierarchical clustering framework of $\Lambda$CDM cosmology, smaller halos form first, with dark matter halos merging to produce progressively larger structures (for a recent review, see \citealt{Baugh2006}; \citealt{Benson2010}). Galaxies form as these halos accrete gas: either shocked as it falls into the halos in a so-called `hot-mode' accretion or along cold filaments of gas which never shock in `cold-mode' accretion (e.g. \citealt{Keres2005}).

Observationally, in a mechanism known as `down-sizing', at high redshifts the most massive galaxies are the most efficient at forming stars \citep{Cowie1996}, with bursts of star formation triggered via mergers of gas-rich disks at high redshift. By $z\sim0$ galaxies with high stellar masses tend to have relatively low gas fractions (GF$\equiv$M$_\text{HI}/$M$_*$) compared to less massive galaxies (e.g. \citealt{Catinella2010}; \citealt{Cortese2011}; \citealt{alfalfa-population}). For an optically selected field galaxy sample with M$_*>10^{10}$ M$_\odot$, typical gas fractions are $0.01<$ GF $<0.1$ (e.g. \citealt{Cortese2011}). At these stellar masses, significantly more gas-rich galaxies are expected to be quite rare, as was found in previous blind \HI\ surveys such as HIPASS \citep{Zwaan2005}.


One of the most surprising results of ALFALFA, the Arecibo Legacy Fast ALFA survey \citep{alfalfa-first} is the detection of a significant population of massive galaxies which are also extremely gas rich (\citealt{Martin2010}; \citealt{alpha.40}). From ALFALFA, we have identified the HIghMass sample: 34 galaxies with high \HI\ mass (M$_\text{HI}\gtrsim10^{10}$M$_\odot$) which are also the most gas-rich ($\text{GF}\gtrsim0.4$) for their stellar masses. Over half of this sample has GF $>1$. A complete definition of the HIghMass sample, along with discussion of the stellar, star formation, and global \HI\ properties can be found in \citet{shanHIghMass}. How do such massive galaxies acquire and maintain substantial reservoirs of gas? We consider two possibilities. 




One possibility is that the gas is recently acquired. This recent gas may be from a `galactic fountain' of gas previously ejected by supernovae raining back down (Shapiro \& Field 1976; Bregman 1980), from accretion of many small companion galaxies, or from accretion of pristine gas along filaments. It is not clear what percentage of the gas comes from each source, although simulations suggest that the majority of gas comes from cooling previously ejected gas, with a smaller---but not negligible---amount coming from cold or hot-mode accretion (\citealt{FraternaliBinney2006}; \citealt{FraternaliBinney2008}; \citealt{Oppenheimer2010}). Regardless of whether such processes dominate, such infalls do not appear to have strong short-scale star formation or stability effects on galaxies but over long times refuel the disk's potential for star formation \citep{Hopkins2013}. Unfortunately, these effects are difficult to find observationally. Direct observation finds gas inflow rates of $\dot{\text{M}}\sim0.2$M$_\odot$\ yr\per{1}\ in local spiral galaxies, a factor of $5-10$ less than is required to sustain star formation \citep{Sancisi2008}. However, rates of both \HI\ and unobserved ionized gas $\dot{\text{M}}\sim1-3$\ M$_\odot$ yr\per{1}\ have been indirectly inferred \citep{FraternaliBinney2008}. 


The second possibility is that the gas is old, but has not been able to form stars. Answers may lie in the HIghMass galaxies' spin parameter $\lambda$ \citep{Peebles1971}, a dimensionless measure of the halo's angular momentum and oft-cited governing mechanism of a galaxy's evolution. Simulations and theory predict that at fixed total mass, galaxies in the local universe with higher spin parameters are bluer, have lower surface brightness, and are more gas-rich (e.g. \citealt{Jimenez1998}; \citealt{MoMaoWhite1998}; \citealt{Boissier2000}; \citealt{Maccio2007}). Unfortunately, the dynamics of the dark matter content of a galaxy are not directly observable, and a number of assumptions are required about the relationship between the spin parameter of the galactic disk ($\lambda_d$) and that of the halo; however, this overall trend appears to be confirmed for inferred spin parameters (e.g. \citealt{hernandez2007}; \citealt{CSHernandez2009}; \citealt{alfalfa-population}). If the HIghMass galaxies have unusually high spin parameters, then this gas may be be unable to efficiently form stars.


This paper presents a detailed view of two HIghMass galaxies, UGC 9037 and UGC 12506. Even among the HIghMass galaxies, these have unusually massive \HI\ reservoirs ($\sim10^{10.5}$ M$_\odot$) and high gas fraction (GF $>1$). They have a low star formation efficiency (SFE$\equiv$SFR$/$M$_\text{gas}\approx10^{-10}$ yr$^{-1}$), typical of the HIghMass sample. We make use of the the Karl G. Jansky Very Large Array (VLA) of the National Radio Astronomy Observatory (NRAO)\footnote{The National Radio Astronomy Observatory is a facility of the National Science Foundation operated under cooperative agreement by Associated Universities, Inc.} to make high-resolution (6\arcsec\ or \about2.9 kpc at 100 Mpc; 7 \kms\ velocity resolution) observations of these two galaxies' \HI\ content and kinematics. This is combined with the Sloan Digital Sky Survey (SDSS) and \Halpha\ observations obtained at Kitt Peak National Observatory (KPNO) to identify regions of active massive star formation activity and to infer the dark matter dynamics of the two galaxies.


Our data reduction methods are found in \S\ref{sec:datareduction}. Details of our analysis methods, including a discussion of how we fit various dark matter halo properties and star formation threshold criteria, are found in \S\ref{sec:analysis}. The results of our observations are discussed in \S\ref{sec:results}. We present our hypotheses for the history of the two galaxies and why they have such high gas masses and gas fractions in \S\ref{sec:discussion}, and summarize in \S\ref{sec:conclusions}.

\section{Observations and Data Reduction}
\label{sec:datareduction}
\noindent
The ALFALFA-derived \HI\ properties, along with optical and UV-derived properties of UGC 9037 and UGC 12506, are summarized in Table \ref{tab:sample}; parameters derived from the resolved VLA maps can be found in Table \ref{tab:vla-properties}. Further discussion of the single dish observations and the derivation of \HI\ properties can be found in \citet{alfalfa-first} and \citet{alpha.40}. Global stellar masses and star formation rates, including corrections for galactic and internal extinction, have been derived for the entire ALFALFA population \citep{alfalfa-population}, following the method of \citet{salim2007}. A detailed discussion, including limitations of this method can be found in \citet{alfalfa-dwarfs}.

\subsection{HI Synthesis Calibration and Data Cubes}
\noindent
\HI\ imaging observations were performed using the VLA from November of 2011 to June of 2012 in the B, C, and D configurations (9.5, 4, and 2 hour observations, respectively) as part of a larger campaign involving 10 HIghMass galaxies. 
Because of technical limitations during the VLA upgrade, observations were performed in multiple spectral windows which were later combined to produce a single 8.5 MHz wide window. Each channel had a native width of 1.7 \kms\ later smoothed to 7 \kms. The full bandwidth of this combined window is $\sim1800$ \kms, enough to both completely encompass each galaxy's \HI\ emission and to fully determine any continuum emission.


Flagging of the visibilities, calibration, and continuum subtraction was performed using standard CASA\footnote{The Common Astronomy Software Applications package is under development by the NRAO.} routines. We produce data cubes using the CASA task \textsc{clean} with a Briggs robustness weighting of 1.0 and the multi-scale clean option. Traditional clean algorithms model the source as a collection of point sources; as we expect the \HI\ to be quite extended, this is a poor assumption. We adopt instead the multi-scale clean method, which models the source as a collection of point sources along with Gaussians of varying sizes (usually the beam width and a few times the size of the beam; \citealt{msclean}). In addition to this being a more physical model of the distribution of the \HI\ emission, multi-scale clean is shown to reduce effects related to over- and under-cleaning the image cubes \citep{msclean-THINGS}. The final noise values of both the data cubes and the flux maps, clean beam sizes, and velocity resolution for the observations are presented in Table \ref{tab:jvla-res}.

\subsection{Maps and Rotation Curves}
\label{sec:mapsrotcurves}
\noindent
Moment maps, rotation curves, and \HI\ density profiles were produced using GIPSY\footnote{The Groningen Image Processing SYstem was developed at the Kapteyn Astronomical Institute.}. Integrated flux maps (also known as moment 0 maps) give a quick overview of the spatial distribution. In order to reduce the impact of noise, we produce them by only summing over the regions in each channel where emission is $>3\sigma$. Velocity fields give a quick overview of the rotational structure of the galaxy; we produce the standard `moment 1' maps, an intensity-weighted velocity average at each sky position. 

For UGC 9037, which is moderately inclined, we decompose the rotational structure seen in the velocity field by fitting a tilted-ring model using \textsc{rotcur} \citep{rotcur}. This model assumes that a galaxy can be decomposed into a series of concentric rings, each of which may have independent inclinations, position angles, rotational velocity, and radial velocities. We choose each ring to have a width of half the beam (3\arcsec). Because the number of free parameters in the fit is very large (4 per ring plus central position and systemic velocity), a simultaneous fit is intractable. We avoid this problem by running \textsc{rotcur} iteratively, keeping different parameters fixed each time. Parameter uncertainties for position angle and inclination are formal errors found by varying the parameters from the values which minimize $\chi^2$. For the rotational and expansion velocities, we use the residual velocity dispersion in each ring to estimate the uncertainty, which tends to dominate over the formal uncertainty. 

At high disk inclinations ($i\gtrsim80\degrees$) as observed in UGC 12506, each line of sight through the galaxy will cross structures at multiple velocities, strongly skewing the line profiles. Rotation curves from velocity fields systematically underestimate the rotational velocities. For such galaxies, the usual method is to fit to the `envelope,' the highest velocities observed at each position observed in a position-velocity (PV) diagram, either by fitting gaussians or tracing a particular isophote (\citealt{SancisiAllen1979}; \citealt{SofueRubin2001}). We use the velocity field and \textsc{rotcur} to estimate a single inclination and position angle for the galaxy, and take a 20\arcsec\ wide slice along the major axis of the galaxy to produce a position-velocity field. At each position along the PV diagram, we extract a spectrum and fit a 3rd order Gauss-Hermite polynomial. The final rotation curve is the position for which the integrated area under the curve is 3.3\% from either the approaching or receding edge. For a gaussian fit (which our fit reduces to for a skew parameter of 0), this is equivalent to the 40\% isophote, a commonly chosen value. Uncertainties are taken to be half the difference between the approaching and receding rotation curves. The actual area chosen has a relatively small effect on the fit: our fits are consistent within errors with those produced by using 1\% to 8\% of the total area.

\subsection{Comparison with ALFALFA Spectra}
\label{sec:spectra}
\begin{figure}
\begin{center}
 \epsscale{0.6}
\plotone{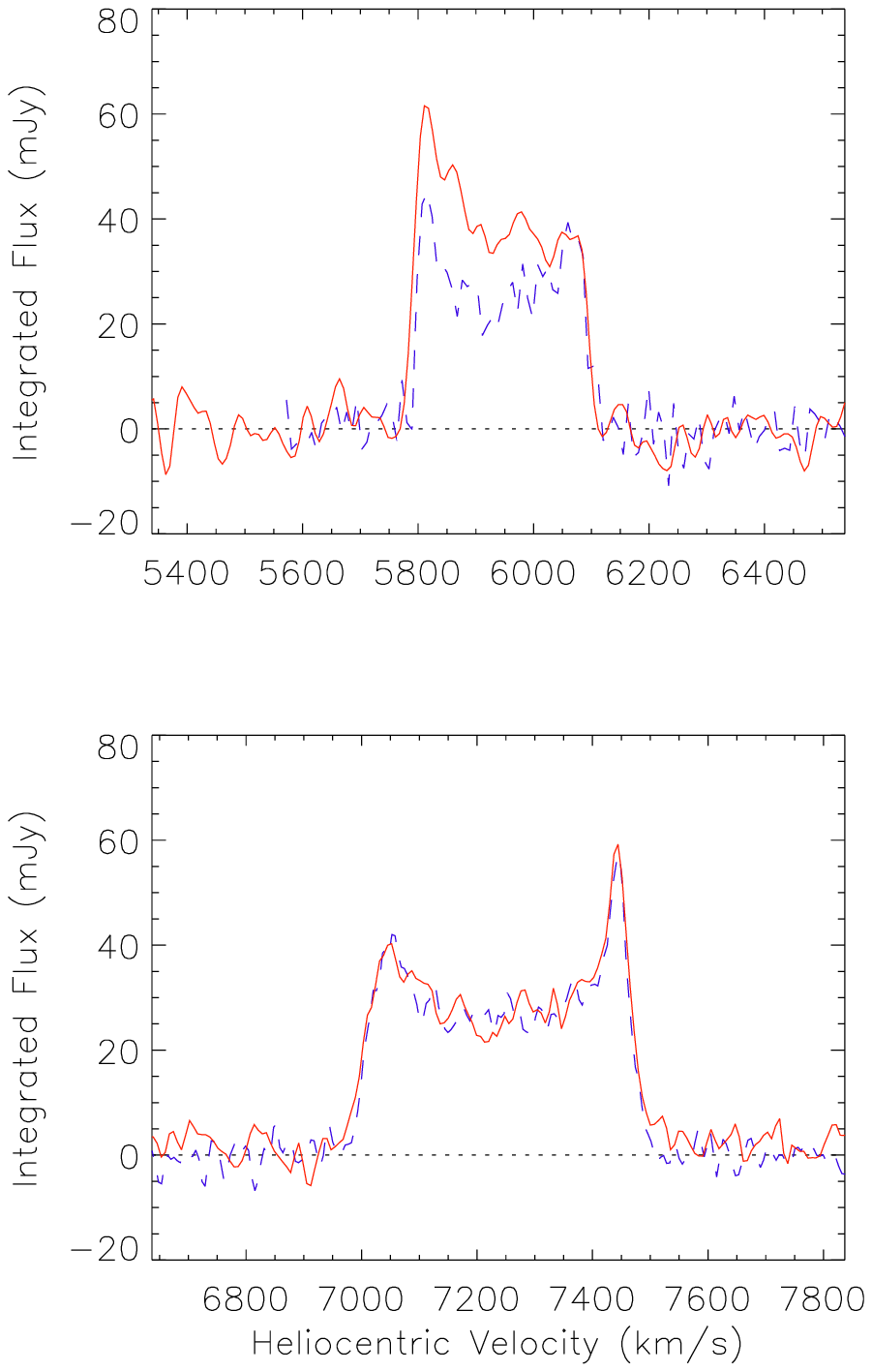}
\end{center}
\caption{
Global spectra of UGC 9037 (top) and UGC 12506 (bottom) as derived from VLA and ALFALFA observations. VLA observations are red solid lines; ALFALFA observations are blue dashed lines. The UGC 12506 spectra match very well, but the VLA spectrum of UGC 9037 shows additional emission in the blueshifted horn compared to ALFALFA.
\label{fig:spectra}}
\end{figure}

\noindent
As global spectra and spectra-derived quantities are the primary data products of ALFALFA, we use these to evaluate the quality of our VLA observations, determine whether significant flux was missed, and double-check our data reduction approach. Figure \ref{fig:spectra} shows our global VLA spectra for UGC 9037 and UGC 12506. The spectra were produced by integrating the data cubes where the flux map indicates that significant ($>3\sigma$) emission exists. The single-dish ALFALFA spectra from the \HI\ archive\footnote{http://arecibo.tc.cornell.edu/hiarchive/alfalfa/default.php} are overplotted. Qualitatively, The VLA spectrum of UGC 12506 closely matches that from ALFALFA. The one for UGC 9037 is less well-matched: the there is significantly more emission in the VLA spectrum compared to the ALFALFA spectrum, particularly in the blueshifted horn of the profile. We do not believe that this asymmetry is real and thus originates either from the higher noise characteristics of the VLA spectrum, 
 or from the uncertainties in the overall flux scale in the VLA observations.

Table \ref{tab:stats} contains the fitted total integrated fluxes, velocity widths, and systemic velocities of the galaxies from the VLA and ALFALFA observations. Discussions of how fluxes and line widths for ALFALFA are derived can be found in \citet{alfalfa-first} and \citet{alpha.40}. We use the same method, except that we correct for instrumental broadening using the relations derived by \citet{hiarchive}, which is not specific to the particular correlator setup of ALFALFA. 
All errors reported are 1-$\sigma$ uncertainties using standard propagation of uncertainties, assuming that the spectral rms is the dominant source of error. A qualitative comparison shows that there are no detectable differences between the spectra for UGC 12506 from ALFALFA and the VLA ($\lesssim 1\sigma$). For UGC 9037, unsurprisingly, the two spectra match more poorly, with the systemic velocity and W50 values being only in reasonable agreement ($\lesssim 2\sigma$). The excess emission observed in UGC 9037 is such that there is overall 9\% more integrated flux in the VLA spectrum of this galaxy compared to the ALFALFA spectrum, a $3\sigma$ difference.

\section{Analysis}
\label{sec:analysis}

\subsection{Mass Models}

\subsubsection{HI and Helium}
\label{sec:gasmodel}
\noindent
\HI\ gas profiles are calculated using the GIPSY task \textsc{ellint}, which integrates the moment 0 map using the tilted ring fit. The gas is assumed to be in an infinitely thin disk, and so face-on gas surface densities are simply a factor of $\cos i$ smaller than the observed values. The derived gas surface densities are largely insensitive to the exact tilted ring fit parameters. For UGC 12506, which is highly inclined, we use the Lucy-Richardson deconvolution (\citealt{Lucy1974}; \citealt{Warmels1988}; GIPSY task \textsc{radial}). This method was specifically designed for galaxies which are only resolved along the major axis. Briefly, this method involves first summing the intensity map along the minor axis, reducing the observed intensities to a one dimensional distribution. A model of axisymmetric, uniform density, coplanar rings is then computed, taking into account that at each line of sight, multiple rings are contributing to the observed intensities. This model is iteratively changed until it matches the observed distribution. The final densities do not require any estimate of the inclination of the galaxy, only a kinematic center and position angle.


We account for Helium by multiplying the \HI\ gas mass by $1.33$ to produce a total atomic gas mass. As the HIghMass galaxies as a whole have abnormally low star formation rates for their \HI\ gas masses \citep{shanHIghMass}, it is unlikely that they significant enough H$_2$ reservoirs to affect the rotation or gas stability in these galaxies. As a rough estimate, assuming a typical massive spiral has an H$_2$ gas depletion timescale of $t_\text{dep}\equiv M_{\text{H}_2}/$ SFR $= \sim10^{9}$ yr (e.g. \citealt{THINGS-sfr}; \citealt{COLD-GASS-ii}), then the H$_2$ makes up only 13\% (for UGC 9037) and 4\% (for UGC 12506) of the total baryonic mass of the galaxies.



\subsubsection{Stellar Mass}
\label{sec:stellarmodel}
\noindent
Compared to our global stellar mass estimates (\citealt{alfalfa-population}; \citealt{alfalfa-dwarfs}), finding from optical photometry a robust measure of how a galaxy's stellar mass distribution is more complex. Among other considerations, it requires an understanding of how the metallicity and extinction change with radius. 
However, our primary interest in a stellar mass density model is in determining its contribution to rotation curves in order to model each galaxy's dark matter content. Since both UGC 9037 and UGC 12506 have M$_\text{HI} > $ M$_*$, the uncertainty on the distribution of stellar mass is less of a factor than correctly estimating the \HI\ plus helium content. As stellar mass is more concentrated towards the center of galaxies compared to the gas, the accuracy of the local stellar mass model is less important than the total stellar mass. To that end, we roughly approximate stellar mass using the following method: elliptical apertures are fit to the optical SDSS image and corrected for Galactic extinction. Internal extinction is ignored because its variation with radius is unknown. Then, the magnitudes are converted into stellar masses using the the M$_*/$L$_i$ and $g-r$ color relations and diet Salpeter initial mass function of \citet{massmodels}. The surface densities are deprojected by rescaling the stellar mass surface densities such that the total integrated mass is equal to the global value. This procedure assumes that the stars are distributed in a infinitely thin disk. We again stress that this method is not intended to give the most accurate distribution of stellar mass, but rather to give a rough estimate of where the bulk of the stellar mass is located.


\subsubsection{Dark Matter}
\label{sec:darkmatter}
\noindent
We use the GIPSY task \textsc{rotmas} to fit dark matter distributions to rotation curves. We model galaxies as composed of a gaseous disk, a stellar disk, and a dark matter halo. The \textsc{rotmod} task calculates the gaseous and stellar contributions to the galaxy rotation, assuming an infinitely thin disk for both. We find that varying models to allow for finite thickness of the stellar disk has little or no effect for these two galaxies. The amplitude of rotation unaccounted for by the stellar and atomic gas components is ascribed to the dark matter: 
\begin{equation}
V_\text{obs}^2 = V_\text{HI}^2 + V_*^2 + V_\text{DM}^2
\end{equation}
We consider two models of the dark matter. The first is the pseudo-isothermal model, which modifies a pure isothermal model (where V is constant at all radii) with a core of constant density.
It is a two-parameter model, including a core scale length $R_C$ and a central density $\rho_0$ which sets the asymptotic velocity. The velocity profile it produces is:
\begin{equation}
\label{eqn:iso-velocity}
V_\text{DM}^\text{ISO}(r) = R_C\sqrt{4\pi G\rho_0\left[1 - \frac{R_C}{r}\arctan\left(\frac{r}{R_C}\right)\right]}
\end{equation}
This asymptotically approaches the constant velocity of a pure isothermal halo, with $V_C \equiv R_C \sqrt{4\pi G\rho_0}$.
The other profile we use is the Navarro-Frenk-White model (NFW; \citealt{NFW}), which is motivated by distributions of collisionless $\Lambda$CDM halos from numerical simulations. It has a velocity profile
\begin{equation}
\label{eqn:nfw-velocity}
 V^\text{NFW}_\text{DM}(r) = V_{200} \sqrt{\frac{1}{x}\frac{\ln(1+cx)-cx/(1+cx)}{\ln(1+c)-c/(1+c)}}
\end{equation}
where $c$ is a parameter giving the central concentration of the halo; only $c>1$ corresponds to physical models. $x\equiv r/R_{200}$, where $R_{200}$ is a characteristic length scale. $V_{200}$ is defined as $V(R_{200})$ and is fixed for given $c$ and $R_{200}$. The NFW profile is thus also a two-parameter family.

\subsubsection{Spin Parameters}
\label{sec:spin-derive}
\noindent
The spin parameter $\lambda$ of the dark matter halo is a dimensionless measure of a halo's angular momentum and is defined as \citep{Peebles1971}:
\begin{equation}
 \lambda = \frac{J\left|E\right|^{1/2}}{GM^{5/2}}
\end{equation}
where $J$, $E$, and $M$ are the halo's total angular momentum, energy, and mass. One unfortunate problem for anyone hoping to fit a dark matter halo's spin parameter is that almost any $\lambda$ can fit fixed NFW halo parameters $c$ and $R_{200}$ or pseudo-isothermal $R_C$ and $\rho_0$. In simulations, $\lambda$ has shown to be only weakly correlated to other halo parameters, with quite large spread (\citealt{Neto2007}; \citealt{Maccio2007}), and so we cannot simply use our rotation curve derived halo properties to constrain the value of $\lambda$. Instead, we use a novel method to estimate the spin parameter by a more direct calculation, finding $J$, $E$, and $M$ in turn.

To simplify computation, it is often assumed that the coupling between the dark and baryonic matter is such that the angular momentum per unit mass $j$ of the dark matter and for the baryons $j_b$ are equal (e.g. \citealt{MoMaoWhite1998}, \citealt{Boissier2000}). The same is true for studies of observed galaxy populations (e.g. \citealt{hernandez2007}, \citealt{CSHernandez2009}, \citealt{alfalfa-population}), which we desire to compare our results with. As such, these calculations are really tracing the so-called modified spin parameter $\lambda^\prime = j_b/j \times \lambda$. For the rest of this paper, when we refer to the spin parameter, or $\lambda$, we are properly talking about the modified spin parameter $\lambda^\prime$. We then calculate the total angular momentum of the dark matter halo:
\begin{equation}
 J = \frac{M}{M_\text{HI}} \sum_i M_{\text{HI},i} V_i r_i
\end{equation}
under the assumption that the gas traces the baryonic angular momentum per unit mass. We could adequately fit tilted rings out to the edge of emission in neither UGC 9037 nor UGC 12506. However, in both galaxies, the rotation curve has either completely flattened (UGC 9037) or is very flat (UGC 12506), and so we assume that a constant velocity continues out to the edge of the \HI\ emission.

We assume that the halos are virialized, and so can estimate their total energy from their kinetic energy. As discussed later in \S\ref{sec:darkmatter-results}, a pseudo-isothermal dark matter model fits both galaxies well. The energy of a pseudo-isothermal dark matter halo can be calculated approximately out to large radii ($r \gg R_C$), where most of the mass and energy reside. Here the pseudo-isothermal model approaches an isothermal halo, which has energy:
\begin{equation}
 \left|E\right| = \left|T+U\right| = T \approx \frac{1}{2} M V^2_C
\end{equation}
where $V_C$ is the maximum circular velocity of the halo and $M$ is the total mass of the dark matter halo. 
Using the NFW fit instead of the pseudo-isothermal changes the estimate in the total energy by only a few percent, and should have little effect on our final results.

We estimate our total halo masses using the results of \citet{baryonmassfunction}. Those authors use abundance matching to relate each observed galaxy's baryonic mass (defined as stellar and \HI) to a simulated halo mass: the  galaxies with the highest baryonic masses are matched with the most massive halos, and so on. In this way, every halo mass can be associated with an average baryonic mass of the galaxy it hosts.

The final dark matter spin parameters calculated for UGC 9037 and UGC 12506 can be found in Table \ref{tab:darkmatter}. We also use the above method to calculate spin parameters of dark matter halos for the THINGS galaxies in \citet{THINGS-rotcur}, using their rotation curves and mass models. The THINGS galaxies act as a reference sample of more `typical' local galaxies than the HIghMass sample. They verify the effectiveness of the above method to calculate spin parameters via comparison to spin parameters empirically-derived from optical properties (e.g. \citealt{hernandez2007}, \citealt{alfalfa-population}) or from simulation (e.g. \citealt{shaw2006}). Discussion of spin parameters and evaluation of this method is presented in \S\ref{sec:darkmatter-results}.


\subsection{Stability and Star Formation Thresholds}
\label{sec:sfr-threshold}
\noindent
We use three theoretical measures to investigate where star formation is expected to occur, in order to compare with our \Halpha\ imaging: the Toomre Q parameter \citep{Toomre1964}, a threshold motivated by galactic shear \citep{Hunter1998}, and a model by \citet{Schaye2004} designed to identify where in a galaxy a cold phase of \HI\ can form. We also consider the gas surface densities at which \HI\ becomes saturated and H$_2$ is observationally found to dominate over \HI \citep{MartinKennicutt2001}. Application of these methods to the densities and rotation curves of UGC 9037 and UGC 12506 is presented in \S\ref{sec:gas-starformation-results}.

An oft-cited criterion for star formation is the Toomre Q parameter. For a thin, differentially rotating disk of pure gas or stars, the dynamics are dominated by pressure, self-gravity, and angular momentum. The gas is stable or unstable to radial (i.e. ring-like) perturbations for certain ranges of surface density and rotational velocities. The condition for stability is \citep{Toomre1964}:
\begin{equation}
 \label{eqn:ToomreQ}
 Q_g = \frac{\sigma_g \kappa}{\pi G \Sigma_g} \geq 1
\end{equation}
where $\sigma_g$ is the velocity dispersion in the gas, $G$ is Newton's gravitational constant, and $\Sigma_g$ is the deprojected surface density of the gas. We take the gas velocity dispersion to be the constant $\sigma_g = 11$ \kms\ found by \citep{THINGS-sfr} from their relatively face-on galaxies. Our galaxies both have $i>60\degrees$, making our own measurements of $\sigma_g$ poor at best. $\kappa$ is the epicyclic frequency:
\begin{equation}
 \kappa^2 = \frac{2\Omega}{r}\frac{d}{dr}\left(r^2\Omega\right)
\end{equation}
where $r$ is the distance from the center of the galaxy and $\Omega(r)$ is the angular speed at radius $r$.

Beyond just gas, one can define analogous $Q_*$, $Q_{\text{H}_2}$ and so on for the various components of the disk. Many prescriptions exist for combining these various $Q_i$ parameters into one effective $Q$. Perhaps the best known is the so-called Wang-Silk approximation, where the \textit{instabilities} of the phases add linearly, or $Q^{-1} = Q_g^{-1} + Q_*^{-1}$ \citep{WangSilk1994}. More rigorous prescriptions are given by \citet{Rafikov2001}, who take into account the wavenumber of such perturbations, as well as \citet{RomeoWiegert2011} and refinements in \citet{RomeoFalstad2013} which weight each of the $Q_i$ based on both the vertical and radial velocity dispersion of each component of the galactic disk. We have found, for each of these three prescriptions, that the $Q_g$ term dominates for our galaxies, and so we approximate $Q \approx Q_g$. In particular, we rearrange Equation (\ref{eqn:ToomreQ}) to express the instability as a critical density above which the gas is unstable against radial flows \citep{Kennicutt1989}:
\begin{equation}
 \SigmaQ = \alpha_Q \frac{\sigma_g \kappa}{\pi G}
\end{equation}
where $\alpha_Q$ is a dimensionless parameter equal to the average $\left<1/Q\right>$ where star formation occurs; again, $\alpha_Q=1$ in the ideal case. Many tests of this parameter have been done in the literature, with the consensus being that star formation occurs in regions for which $\alpha_Q < 1$. As a few examples, \citet{Kennicutt1989}\footnote{They report that $\alpha_Q = 0.67$, but they have a slightly different definition of \SigmaQ.} found $\alpha_Q=0.63$, \citet{MartinKennicutt2001} found $\alpha_Q=0.69$, though they both assume $\sigma_g=6$ \kms, which more closely matches the dispersion of the cold phase alone. Changing this assumption to our own $\sigma_g=11$ \kms\ yields $\alpha_Q=0.34$ and $\alpha_Q=0.38$. \citet{THINGS-sfr} found $\alpha_Q\approx0.4$ for their massive spiral galaxies, using the same assumptions that we do. We choose $\alpha_Q = 0.4$ as a typical `marginally unstable' threshold as observed in the literature. The deviation of $\alpha_Q$ from unity may be an indication that star formation can occur long before the gas is dense enough to cause axisymmetric inflow of the gas. Alternatively, other parameters, 
such as the presence of stars (e.g. \citealt{JogSolomon1984}, \citealt{JogSolomon1992}) have significant impacts on the stability. This discrepancy may also be due to using global properties, such as the rotation curve and averaged gas densities to predict star formation, which is ultimately a local process (e.g. \citealt{Krumholz2012}).

\citet{Hunter1998} take a different approach to star formation from large-scale collapse in the gaseous disk. They examine the case where the \HI\ clouds lose angular momentum along magnetic field lines. In this case, the dominant effect preventing collapse becomes the shear in the disk. This critical density can be expressed as:
\begin{equation}
 \SigmaA = \alpha_A \frac{\sigma_g A}{\pi G}
\end{equation}
where $\alpha_A$ is an uncertain dimensionless parameter, similar to $\alpha_Q$. $A$ is Oort's A constant, defined as:
\begin{equation}
 A = -0.5 r \frac{d}{dr}\Omega(r)
\end{equation}
\citet{Hunter1998} chose $\alpha_A = 2.5$ based on a rough argument that perturbations must grow by a factor of 100 to be significant, while \citet{Schaye2004} implies $\alpha_A = 1$ at the onset of instability. In a sample of 30 nearby spirals, \citet{MartinKennicutt2001} suggest $\alpha_A = 2.0$. We use $\alpha_A=2.5$ as a baseline for the maximum required surface density. 

A third approach to star formation, taken by \citet{Schaye2004}, is to instead consider the regions in the disk where a cold phase of gas and the production of \Htwo\ can occur. This cold phase will have a lower velocity dispersion, 3 \kms$<\sigma_\text{cold}<6$ \kms, compared to that in the dominant warm phase, $\sigma_\text{gas}=11$ \kms, and such a phase has a correspondingly lower \SigmaQ\ and \SigmaA. \citet{Schaye2004} derives criteria under which a cold phase can form. These criteria can be expressed as:
\begin{equation}
 \SigmaS \approx 6.1 \frac{\Sigma_\text{gas}}{\Sigma_\text{gas}+\Sigma_*}\text{M}_\odot\text{ pc}^{-2}
\end{equation}
for an assumed constant metallicity of $0.1$ $Z_\odot$ and interstellar flux of ionizing photons $10^6$ cm\per{2}\ s\per{1}.

Finally, it has long been observed that \HI\ surface densities of $\sim9-10$ \solarmassespersquareparsec\ are `saturated,' (e.g. \citealt{MartinKennicutt2001}, \citealt{Bigiel2008}, \citealt{THINGS-sfr}) in the local universe, and densities exceeding this value are rarely observed; excess gas exists in the molecular phase. This does not appear to be a threshold for star formation, only a location where the star formation efficiency becomes roughly constant \citep{THINGS-sfr}.

\section{Results}
\label{sec:results}

\subsection{UGC 9037: Images and Rotation Curve}
\label{sec:results-UGC9037}
\begin{figure}
\begin{center}
 \epsscale{0.35}
\plotone{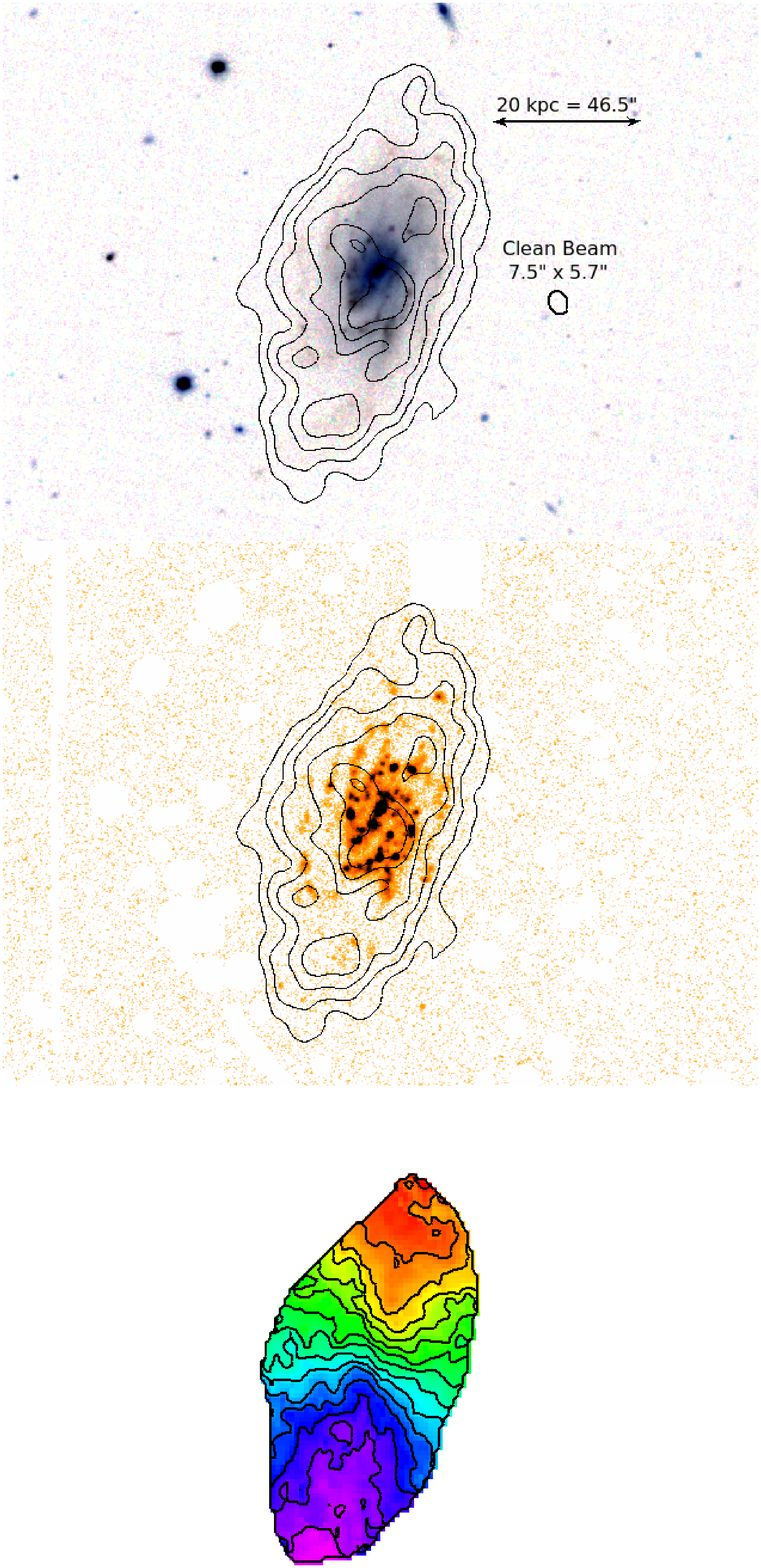}
\end{center}
 \caption{Images of UGC 9037, which has an inclination of $\sim60^\circ$. (top) \HI\ intensity map. Contours begin at 4$\sigma$ ($N_\text{HI}=10^{21}$ cm\per{2}), and increase by $2\sigma$ at each additional contour. The contours are overlaid on a false-color SDSS image ($g$, $r$, and $i$ bands) with each color channel adjusted for maximum contrast. (middle) \Halpha\ image of UGC 9037 from \citet{shanHIghMass}, with \HI\ density contours overlaid. \HI\ emission is tightly peaked near the center of the galaxy and does not extend significantly beyond the stellar disk; the \Halpha\ shows significant star formation in the spiral arms which smoothly decreases with radius. (bottom) Moment 1 velocity field. Isovelocity contours are stepped at 20 \kms.\label{fig:images-UGC9037}}
\end{figure}

\noindent
Figure \ref{fig:images-UGC9037} presents images of UGC 9037 \citep{shanHIghMass} in several different bands. At the Hubble flow distance of $\sim89$ Mpc (assuming H$_0=70 $\kms\ Mpc\per{1}), each arcsecond corresponds to 0.42 kpc, and so with a clean beam of $\sim6$\arcsec, the VLA observations have a resolution of roughly 2.5 kpc. The top left panel is a SDSS optical image, with contours from our integrated flux map overlaid. The contours begin at 2$\sigma$ significance ($N_\text{HI}=10^{21}$ cm\per{2}), and increase by $2\sigma$ at each additional contour. On the scale of the beam, significant substructure is clearly visible. The peak of central \HI\ emission is slightly displaced from the optical center of the galaxy, and the spiral arm features to the south are clearly visible in \HI. A weak bar is optically visible, but we do not have the resolution in \HI\ to determine any relationship between the stellar population and \HI\ there.

One interesting feature is that even though UGC 9037 is gas dominated (GF$>1$), its \HI\ does not appear to be more extended beyond its optical radius compared to less gas-rich objects of similar stellar mass. The ``Bluedisks'' sample examined by \citet{Bluedisks} includes galaxies of similar stellar mass but $\text{GF}\sim0.3$. For that sample, typical ratios of \HI\ radius ($R_\text{HI}$; where the deprojected $\Sigma_\text{HI}=1$ \solarmassespersquareparsec) to optical radius ($R_{25}$; where the B-band surface brightness reaches 25 mag \arcsec\per{2}) are $0.1<\log(R_\text{HI}/R_{25})<0.4$ at the same stellar mass. UGC 9037 has a value in the center of that range, $\log (R_\text{HI}/R_{25})=0.26$.

If we consider only the radius of the \HI\ itself, we also find that UGC 9037 is not extended compared to other galaxies of similar \HI\ mass. \citet{BroeilsRhee1997} found a very strong ($r=0.98$) linear relationship between $\log$M$_\text{HI}$ and $\log R_\text{HI}$, which was observed to extend to the gas-rich more massive galaxies of the `Bluedisks' sample (\citealt{Bluedisks}; see Equation 3 of \citealt{BroeilsRhee1997}):
\begin{equation}
  \label{eqn:Broeils}
  \log R_\text{HI} = 0.51 \log \text{M}_\text{HI} - 3.63
\end{equation}
For UGC 9037, the predicted \HI\ radius from the VLA maps is $44\pm4$ kpc, in good agreement with the observed $42.1\pm0.7$ kpc. From the optical and \HI\ radii alone, it would appear that UGC 9037 is a `normal' spiral galaxy which has simply been scaled up to a high \HI\ mass and gas fraction. 

The middle panel of Figure \ref{fig:images-UGC9037} is an \Halpha\ image of UGC 9037 with the same \HI\ contours overlaid. Star formation is confined to the spiral arms and bar and decreases smoothly out to the edge of detectable emission at \about25 kpc. The bottom panel is the moment 1 velocity field, with contours spaced by 20 \kms. The field shows smooth, well-defined velocities across the extent of \HI\ emission, but some deviations from perfect rotational motion can be seen: there is a twist in the isovelocity contours near the galaxy center along the major axis.

\begin{figure}
\begin{center}
 \epsscale{1.0}
\plotone{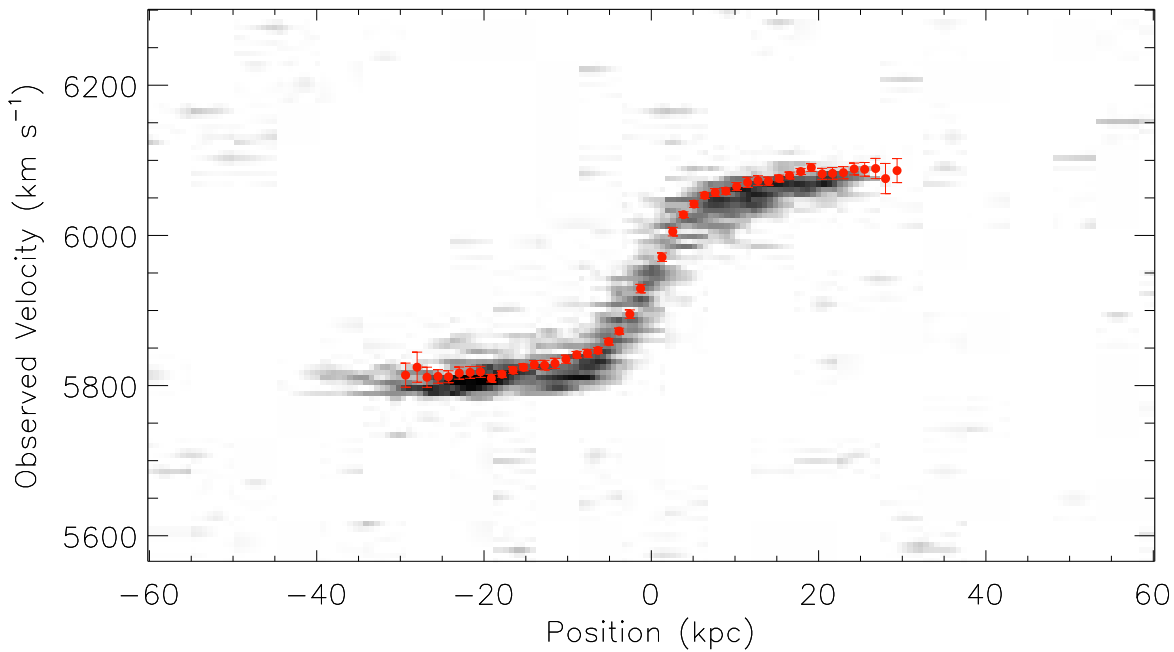}
\end{center}
\caption{Position-Velocity (PV) diagram of UGC 9037, depicting a slice taken through the center of the data cube at the average position angle of the galaxy (-11\degrees). Velocities are observed velocities on the sky, not corrected for inclination. Negative positions are south of the galaxy center, positive positions are to the north. Overplotted points are the fitted rotation curve (see Figure \ref{fig:rotcur-UGC9037}), projected onto the sky. The PV diagram is consistend with a rapidly rising rotation curve which quickly reaches the maximum rotation velocity; some evidence of decline is seen on the approaching side.\label{fig:PV-UGC9037}}
\end{figure}

Figure \ref{fig:PV-UGC9037} shows a position-velocity map, made by taking a slice through the data cube along the average position angle of the galaxy (11\degrees\ west of north), 20\arcsec\ wide, through the center of the cube. The velocity structure of UGC 9037 is very clean: at every position along the galaxy, there is a single, narrow range over which \HI\ emission is present.  The only asymmetry in the galaxy is the mismatch of the radial extent of the approaching and receding sides of the galaxy also apparent in Figure \ref{fig:images-UGC9037}. While the rotation of the receding (north) side of the galaxy can be traced only to 25 kpc, the approaching (south) side is more extended, with emission still easily visible to $\gtrsim40$ kpc. The derived rotation curve is over-plotted on the diagram, projected onto the sky, and matches very well, but the asymmetric nature of the emission is not captured in the rotation curve, which is assumed to be symmetric.

\begin{figure}
\begin{center}
 \epsscale{1.0}
\plotone{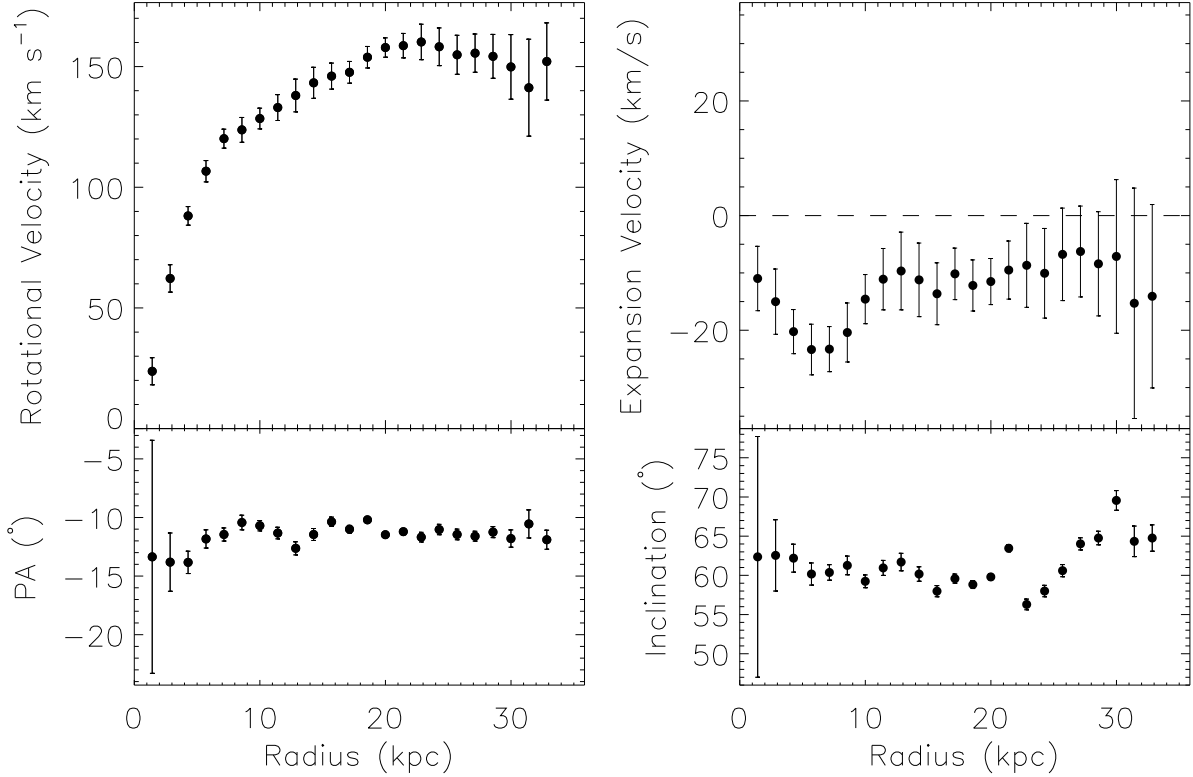}
\end{center}
\caption{Decomposition of UGC 9037 into tilted rings fit to the velocity field. Subgraphs are (top left) rotational velocity, corrected for inclination; (top right) expansion, or radial velocity of the rings; (lower left) position angle of rings, measured east from north (all fitted angles are negative, and thus the semimajor axis is slightly west of north); (lower right) inclination of rings; 0\degrees\ is face on. The rotational velocity rises quickly to a maximum rotational velocity of \about160 \kms\ after 20 kpc, qualitatively matching the PV diagram (see Figure \ref{fig:PV-UGC9037}). Across the disk, there are clear indications of noncircular gas motions: the gaseous disk is rapidly rotating, but also showing non-circular velocities. The disk geometry is quite flat, with the position angle and inclination varying by only 5-10\degrees\ across the entire galaxy.\label{fig:rotcur-UGC9037}}
\end{figure}

The decomposition of UGC 9037 into tilted rings appears in Figure \ref{fig:rotcur-UGC9037}. All four panels are plotted as a function of deprojected radius to the fitted rotational center. The black filled points and solid lines are for the best-fit model. In the top left panel is the rotational velocity. The rotation curve of UGC 9037 rises quickly, and by 20 kpc has reached its maximum of 160 \kms, then shows signs of decline. The top right panel presents the expansion velocities of the rings. These velocities are negative in the \textsc{rotcur} model, but without knowing which side of the disk is closest, it is not possible to determine whether these velocities represent inflow or outflow. There are non-circular velocities at all radii, although only within 20 kpc does each individual point become significant. The average non-circular velocity is $-14.4\pm1.2$ \kms, approximately 9\% of the maximum rotation speed of $160.2\pm7.3$ \kms.

Are these non-circular velocities real and significant? One problem with fitting tilted ring model rotation curves is that disk geometry is degenerate with non-circular velocities: one can, by varying the position angle and inclination of the disk, allow for any choice of `expansion velocity' (e.g. \citealt{Schoenmakers1997}; \citealt{SellwoodSanchez2010}). However, it is well known that the inner regions of real galaxy disks are very flat, and our best-fitting model is also quite flat. The position angle (measured east of north; lower left) and inclination (lower right) only vary by a few degrees over the entire disk, although outside of 20 kpc, the disk shows some slight bending ($\Delta i \sim 10\degrees$). Forcing the non-circular velocities to zero requires the inner disk to twist quite strongly ($\Delta$ PA $\sim 15\degrees$), strongly counter-indicating such a model. The inner 10 kpc of our model disk does appear to twist slightly ($\Delta$ PA $\sim5\degrees$), although such changes are within our uncertainties. This small twisting could be an indication of the fit following the bar and spiral arms apparent in the optical image and \HI\ intensity map. Such a twisting is observed in resolved studies of nearby galaxies (e.g., \citealt{Chemin2009} and \citealt{Corbelli2010}). However, unlike those authors' findings, fixing the position angle in the inner disk does not remove the effect.

\begin{figure}
\begin{center}
 \epsscale{0.9}
\plotone{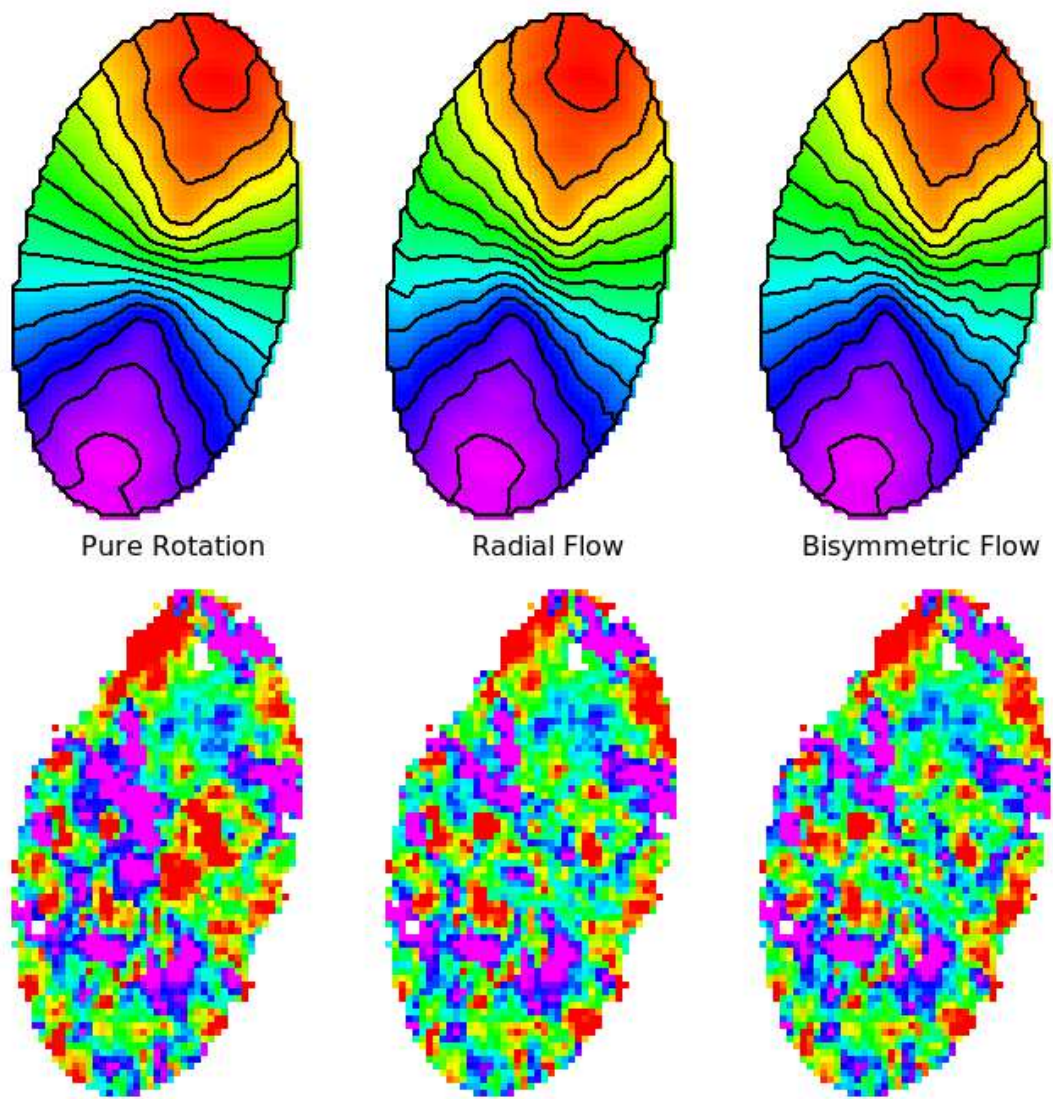}
\end{center}
\caption{(top) Model velocity fields of UGC 9037 created with DiskFit, with solid lines every 20 \kms; (bottom) residuals with respect to the observed velocity field (see Figure \ref{fig:images-UGC9037}, bottom panel). The residual color scale ranges from $-10$ \kms (purple) to $+10$ \kms (red). (left) Model with pure rotational velocities, (middle) model allowing for symmetric radial flows, and (right) model allowing for bisymmetric (bar-like) flows. The residuals of the pure rotation model are correlated, and have opposite signs in the east and west halves of the inner disk, showing this model to be a poor fit to the data. The residuals for the radial and bisymmetric flow models are of nearly equal quality.
\label{fig:diskfit}}
\end{figure}

Second, we note that while the tilted ring model implemented by \textsc{rotcur} can probe some non-circular motions, it is primarily designed to model flat disks dominated by rotational motions with possible warping at large radii; the only non-circular motions it can account for are axisymmetric radial flows. To supplement the \textsc{rotcur} results, we use the DiskFit package (\citealt{SpekkensSellwood2007}; \citealt{SellwoodSanchez2010}; \citealt{KuziodeNaray2012}), which fits more realistic flat disk models to the observed data. We fit three models using DiskFit: a pure rotational model, a model which allows for radial flows, and one which models bisymmetric (bar-like) flows. The flat disk geometry is allowed to vary in the models, and in all cases we find values that are consistent with the average values obtained by \textsc{rotcur}. Figure \ref{fig:diskfit} presents the resulting model velocity fields (top) and residuals compared to the moment 1 map (bottom), for the pure rotational (left), radial flow (middle) and bisymmetric flow (right) models. The model velocity fields have isovelocity contours every 20 \kms, and the color bar for the residuals spans -10 \kms\ to 10 \kms. As with \textsc{rotcur}, a pure rotational model fits the inner disk poorly: the residuals show opposite systematic trends in the east and west halves of the inner disk. By eye, the radial and bisymmetric flow models appear to have nearly identical residuals, and both reproduce the inner velocity field twist (Figure \ref{fig:images-UGC9037}, bottom panel). Both non-circular flow models present plausible representations of the data: it is not possible to distinguish between the two using the quality of the fit \citep{SpekkensSellwood2007}. However, the presence of a weak bar in UGC 9037 (Figure \ref{fig:images-UGC9037}, top panel) suggests that bar-like flows are the most likely explanation for the observed non-circular motions, although streaming motions toward the galaxy center may also be consistent with the picture that UGC 9037 is on the verge of undergoing intense star formation. All three models show some systematic trends in their residuals at large radii; this may be related to the slight warping seen at large radii in the \textsc{rotcur} rotation curve, which has not been taken account in DiskFit.



\subsection{UGC 12506: Images and Rotation Curve}
\label{sec:results-UGC12506}
\begin{figure}
\begin{center}
 \epsscale{0.4}
\plotone{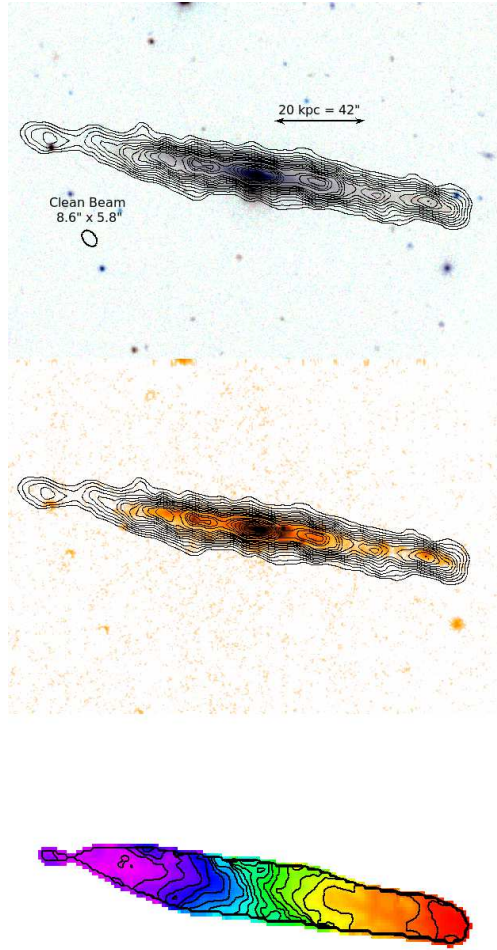}
\end{center}
 \caption{Images of UGC 12506, which has an inclination of $\sim86^\circ$. Panel sequence is analogous to that in Figure \ref{fig:images-UGC9037}; the 4$\sigma$ here is $N_\text{HI}=8.4\times10^{20}$ cm\per{2}. Here the \HI\ is extended beyond the stellar disk ($\log R_\text{HI}/R_{25} = 0.26$), a value typical for galaxies of similar stellar mass and lower gas fraction (see \citealt{Bluedisks}). The \Halpha\ shows that star formation occurs across the entire optical disk. \label{fig:images-UGC12506}}
\end{figure}

\noindent
In the same format as Figure \ref{fig:images-UGC9037}, Figure \ref{fig:images-UGC12506} presents images of UGC 12506. At the Hubble flow distance of 98 Mpc (assuming H$_0=70$\kms\ Mpc\per{1}), each arcsecond corresponds to 0.48 kpc, and so with a clean beam of $\sim6$\arcsec, the VLA observations have a resolution of roughly 2.9 kpc. The top panel shows \HI\ contours overlaid on a SDSS optical image of UGC 12506. The contours begin at 4$\sigma$ significance ($N_\text{HI}=8.0\times10^{20}$ cm\per{2}), and increase by 2$\sigma$ at each additional contour. UGC 12506 is very large galaxy. Its optical radius is 40 kpc, and the \HI\ can be traced beyond 60 kpc. Like UGC 9037, it has a typical, but lower, $\log (R_\text{HI}/R_{25}) = 0.16$. Similarly, the predicted \HI\ radius for a galaxy of its \HI\ mass (from Equation \ref{eqn:Broeils}) is $56\pm4$ kpc, in agreement with the observed $58\pm2$ kpc. From the optical and \HI\ radii, the galaxy appears fairly typical for its stellar and \HI\ masses.  

In the optical image there are two interlopers, neither of which is connected with UGC 12506. To the east, there is a star nearly coincident with a small overdensity of \HI, and to the south, intersecting with UGC 12506 is a galaxy at higher redshift.


The middle panel of Figure \ref{fig:images-UGC12506} shows our H$\alpha$ image of UGC 12506. UGC 12506 shows star formation smoothly spread throughout the disk. The eastern overdensity of \HI\ shows no significant star formation. The bottom panel of Figure \ref{fig:images-UGC12506} is the velocity field. Isovelocity contours are spaced every 20 \kms. The galaxy shows ordered rotation out to the edge of where emission is detectable. The velocity field of UGC 12506 is more complex compared to the smooth UGC 9037. This is an effect of the high inclination of the disk; we do not believe that the closed contours indicate a real decrease in the rotation curve. As mentioned in \S\ref{sec:mapsrotcurves}, the inclination of UGC 12506 is high ($i = 86\degrees$), causing the velocity field to be a poor indicator of the rotational velocity of the galaxy: a position-velocity diagram is required.

\begin{figure}
\begin{center}
 \epsscale{1.0}
\plotone{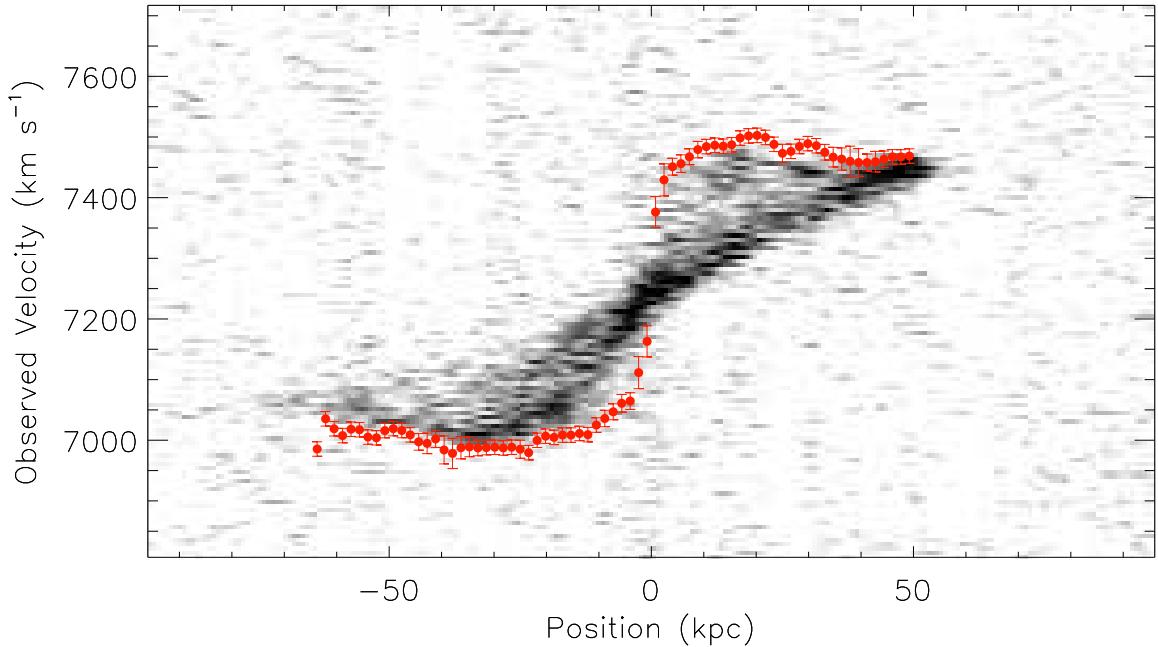}
\end{center}
\caption{Position-Velocity diagram of UGC 12506, depicting a slice taken through the center of the data cube at the position angle of the entire galaxy (-99.7\degrees). Velocities are observed velocities on the sky, not corrected for inclination. Negative positions are east of the galaxy center, positive positions are to the west. Overplotted points are the fitted rotation curve, allowing for the significant asymmetry in the approaching and receding arms, as fit to the PV diagram. UGC 12506's PV diagram shows much more complex structure as emission is not confined to a narrow range of velocities at each position. This is due to the high inclination of UGC 12506 ($i\sim86^\circ$): at every off-center position a range of projected velocities are compatible with a single rotational velocity. 
\label{fig:PV-UGC12506}}
\end{figure}

Figure \ref{fig:PV-UGC12506} shows the position-velocity diagram for UGC 12506, produced in the same fashion as Figure \ref{fig:images-UGC9037}, except that the major axis is along 99.7\degrees\ west of north. The galaxy's high ($\gtrsim80^\circ$) inclination combined with the finite \HI\ beam size means that there is a broad range of velocities present at every position along the major axis, rather than a single velocity. The envelope fitting method described in \S\ref{sec:mapsrotcurves} smoothly traces the maximum observed velocities. There is some asymmetry in the approaching (eastern) and receding (western) sides of the galaxy in both the shape and length of the arms. The receding side is detectable only to $\sim50$ kpc from the galaxy center, while the approaching side is detectable to $\sim70$ kpc. 

\begin{figure}
\begin{center}
 \epsscale{1.0}
\plotone{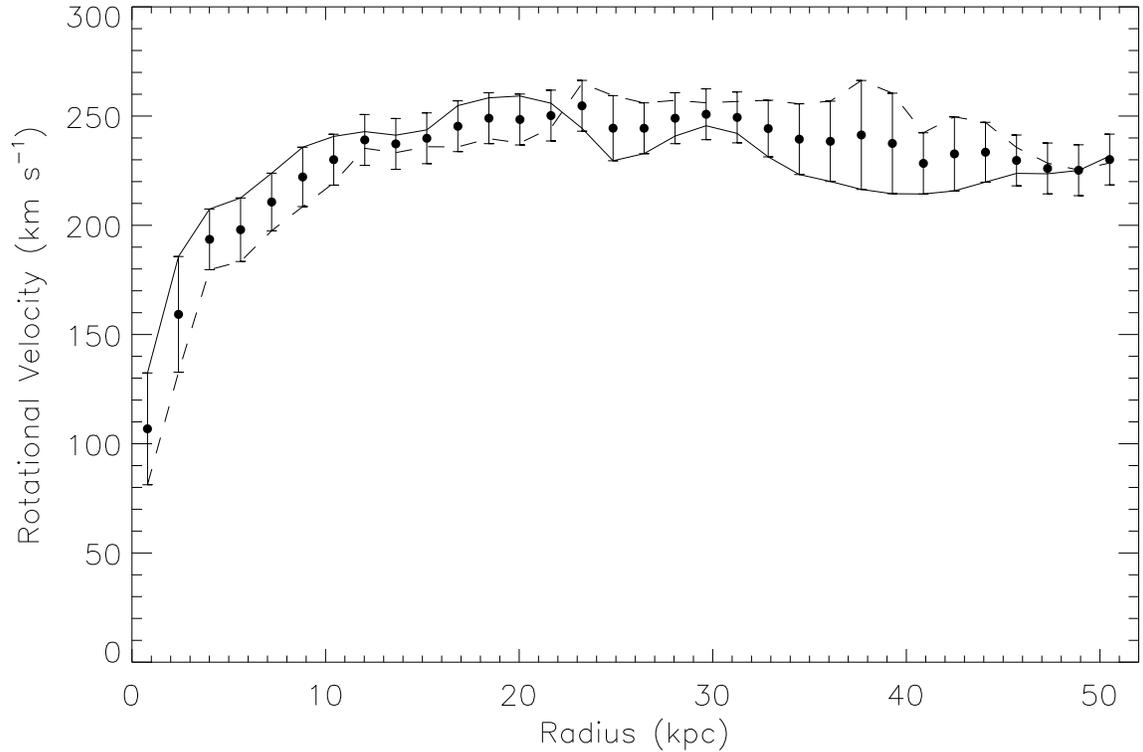}
\end{center}
\caption{Rotation curve of UGC 12506, the average of the approaching and receding sides, as observed in Figure \ref{fig:PV-UGC12506}, and corrected for inclination. The lines represent the rotation curve on the receding (solid) and approaching (dashed) sides of the galaxy, while the points are the average of the two. The entire galaxy is assumed to have a single inclination ($i=86.4\degrees$) and positional angle ($PA=99.7\degrees$ west of north). Like UGC 9037, the rotation curve rises rapidly, but unlike UGC 9037 it shows signs of decline at large radii. 
\label{fig:rotcur-UGC12506}}
\end{figure}

UGC 12506's rotation curve, calculated from the PV diagram, is presented in Figure \ref{fig:rotcur-UGC12506}. The two lines show the receding (solid) and approaching (dashed) arms of the galaxy. The rotation curve rises quickly and reaches a maximum velocity of $\sim250$\kms.


\subsection{Gas Stability and Star Formation}
\label{sec:gas-starformation-results}
\begin{figure}
\begin{center}
 \epsscale{0.73}
\plotone{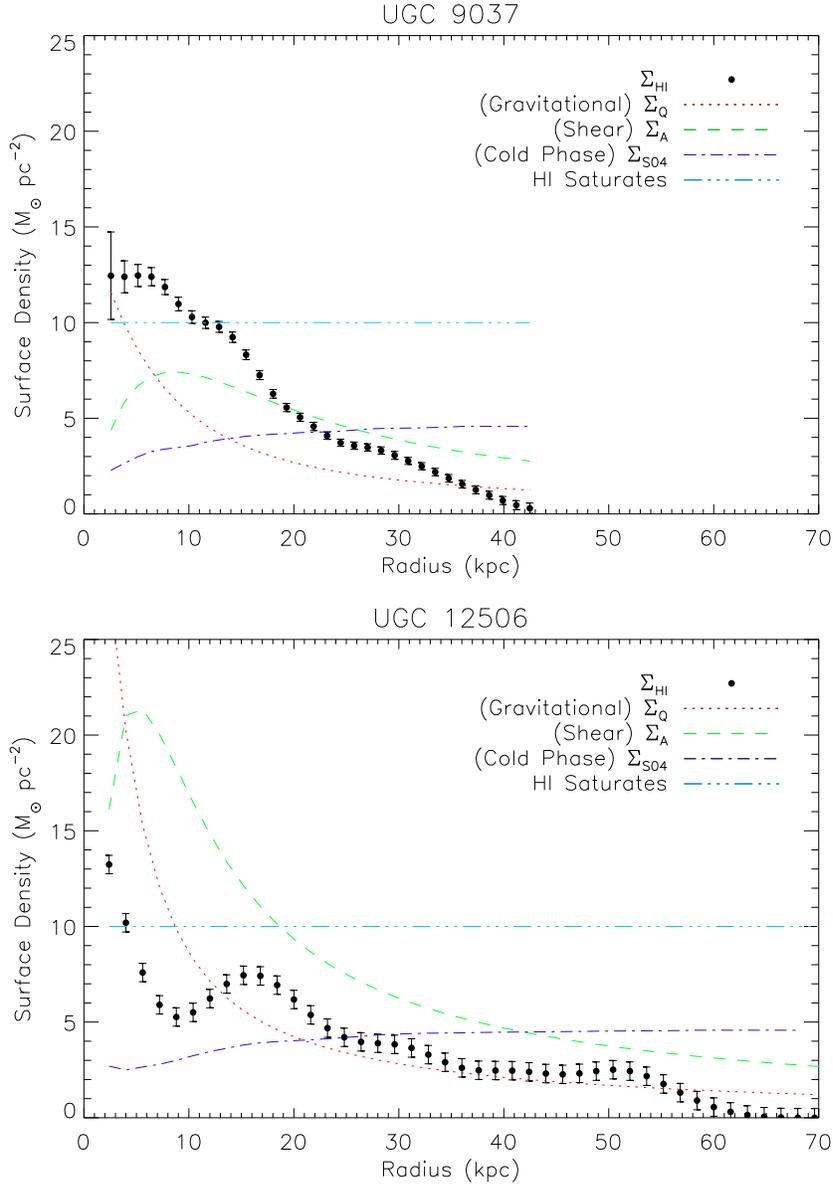}
\end{center}
\caption{\HI\ surface densities as a function of radius for both UGC 9037 (top panel) and UGC 12506 (bottom), on identical axes (thick black lines). Also plotted are critical densities for: ring-like gravitational collapse ($\Sigma_Q$; red dotted lines), collapse inhibited by shear ($\Sigma_A$; green dashed lines), for a cold phase to form ($\Sigma_\text{S04}$; purple dashed-dotted lines), and where \HI\ is seen to saturate (\HI\ saturates; cyan dashed triple-dotted lines). The \HI\ in UGC 9037 is more centrally peaked, with surface densities $>10$ \solarmassespersquareparsec, while the \HI\ in UGC 12506 is more extended, out to 60 kpc, with typical surface densities of $1-4$ M$_\odot$.\label{fig:density}}
\end{figure}

\noindent
Figure \ref{fig:density} shows the deprojected \HI\ surface densities (thick black lines) of UGC 9037 (top) and UGC 12506 (bottom) as a function of radius from the center of the galaxy. Also plotted are the thresholds for axisymmetric gravitational instability (\SigmaQ; dotted red lines), shear instability (\SigmaA; dashed green lines), for a cold phase to form (\SigmaS; dash-dotted purple lines), and where \HI\ is observed to saturate ($\sim10$ \solarmassespersquareparsec; cyan dashed triple-dotted line).

The central \HI\ density is quite high in UGC 9037, reaching almost 14 \solarmassespersquareparsec, larger than the 10 \solarmassespersquareparsec\ where \HI\ is observed to saturate. In the THINGS sample, high surface densities are only observed in NGC 3077 (whose gas is highly disturbed and asymmetric) and NGC 4214 (which is an irregular galaxy). Compared to UGC 9037, both have over an order of magnitude less mass in stars and \HI\ ($\text{HI}_\text{HI}\sim\text{M}_*\sim9$). In addition, over the entire disk, the \HI\ density exceeds our $\Sigma_Q$ criterion, indicating that the gas is marginally unstable. In this interpretation, the non-circular motions detected in the velocity field may represent inflowing gas to fuel the imminent star formation. A detailed investigation of higher-resolution kinematics, such as those afforded by CO interferometry, as well as a determination of the near side of UGC 9037's disk may help distinguish radial inflows from other interpretations of the observed non-circular motions.


Despite having a slightly higher overall \HI\ mass compared to UGC 9037, UGC 12506 has a low density (typically $1-5$ \solarmassespersquareparsec) \HI\ disk out to 60 kpc. Only in the very center of the galaxy ($<5$ kpc) do we expect \Htwo\ to be an important component of the interstellar medium. UGC 12506 lies just at the $\Sigma_\text{HI}=\Sigma_Q$ line except in the inner \about15 kpc, where $\Sigma_\text{HI}$ becomes strictly less than $\Sigma_Q$. However, the \Halpha\ imaging does not mirror this: we see healthy star formation throughout the disk, including near the galaxy's center. If \Htwo\ is a significant near the center of the galaxy, it may make up this gap.

Both galaxies have a roughly exponential \HI\ profile, unusual for galaxies in the local universe. Generally, a flat or decreasing density in the inner disk with $\Sigma_\text{HI}$ stagnating near 10 \solarmassespersquareparsec\ is observed (e.g. \citealt{THINGS}; \citealt{THINGS-sfr}). UGC 9037 only begins to flatten in the very inner disk, at approximately 13 \solarmassespersquareparsec\ of \HI. As UGC 12506 barely reaches 10 \solarmassespersquareparsec\ of \HI\ near its center, and thus could not show a flat profile over any significant range of radii given the resolution of our observations, it is difficult to draw conclusions about its profile. However, the combination of its very high \HI\ mass and low surface densities are very unusual.

For neither galaxy does the shear criterion appear to be a significant predictor of star formation, as for both galaxies $\Sigma_\text{HI} < \Sigma_A$ across essentially the entire disk despite clear evidence of star formation across the entire disk. Similarly, for both galaxies, the predictions of the model of \citet{Schaye2004} suggest that a cold phase can form only in the inner $10-20$ kpc of each galaxy.

\subsection{Dark Matter and Spin Parameters}
\label{sec:darkmatter-results}
\noindent
The parameters obtained from fits of both NFW and ISO halos to our galaxies' rotation curves appear in Table \ref{tab:darkmatter}. Both galaxies are well-fit by both halo models ($\chi^2/\text{dof}\lesssim1$). An F-test statistic indicates that UGC 9037 shows weak evidence of favoring the ISO model ($\chi^2_\text{ISO} = 0.65; \chi^2_\text{NFW} = 1.29$; $p = 0.07$), but is not conclusive. For UGC 12506, there is a very significant preference for the NFW model ($\chi^2_\text{ISO} = 0.54; \chi^2_\text{NFW} = 0.21$; $p = 0.004$). As UGC 12506 is a low surface brightness galaxy \citep{shanHIghMass}, this result is surprising. In general, LSB galaxies exhibit slowly rising rotation curves and cannot be fit by NFW profiles (e.g. \citealt{Moore1994}, \citealt{McGaugh1998}), while UGC 12506 exhibits neither of these qualities.

\begin{figure}
\begin{center}
 \epsscale{0.8}
\plotone{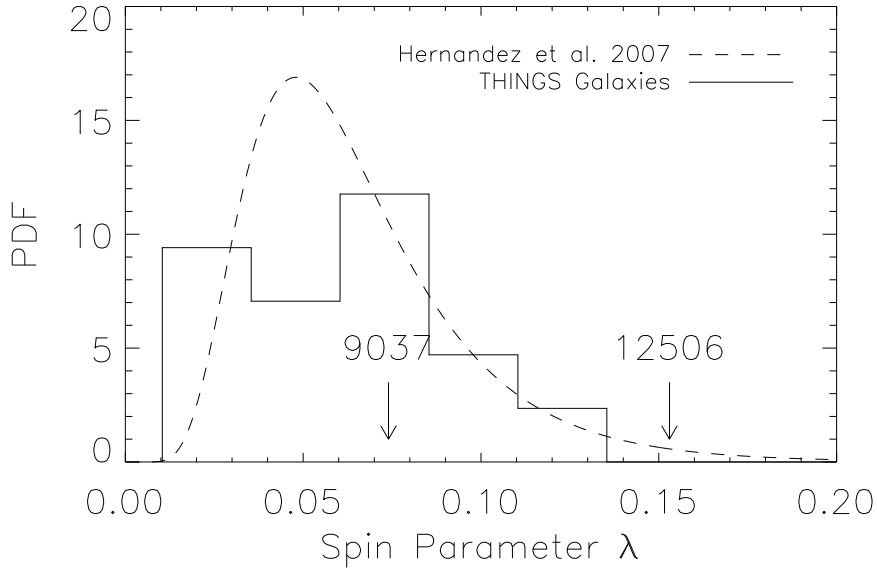}
\end{center}
\caption{Distribution of observed spin parameters of galaxies. The solid histogram shows spin parameters of the THINGS galaxies. The dashed line is the best-fit lognormal distribution of SDSS galaxies by \citet{hernandez2007}. The two very roughly occupy the same range of values. They may be from the same distribution, but with the small number of galaxies in the THINGS sample we cannot tell. Arrows show the calculated spin parameters of UGC 9037 (0.07) and UGC 12506 (0.15). Measured against either standard, UGC 9037 has a typical spin parameter, while UGC 12506's is quite high.
\label{fig:lambda}}
\end{figure}

As previously discussed, simulations and semi-analytic models suggest a correlation between gas richness and high spin parameter as suggested observationally by \citet{alfalfa-population}. Figure \ref{fig:lambda} shows a probability density function of spin parameters of THINGS galaxies calculated according to the method in \S\ref{sec:spin-derive}; the histogram is normalized so that the total area is unity. The range of spin parameters ($0.01<\lambda<0.10$) that we calculate for the THINGS galaxies matches up with `typical' values of spin parameters derived from observation and simulation (\citealt{shaw2006}; \citealt{Maccio2007}; \citealt{hernandez2007}). Also plotted in the figure as a dashed line is the best-fit lognormal distribution of the sample of \citet{hernandez2007}. The two are similar in a very rough sense. The disagreement may be due to the differences between our method and that of \citet{hernandez2007} for calculating spin parameters, but given the small number statistics and uncertainties in the calculation of $\lambda$, we hesitate to read further into this.

The spin parameters for our two galaxies, UGC 9037 and UGC 12506 are plotted in Figure \ref{fig:lambda} as arrows. UGC 9037 has a spin parameter of $\lambda=0.07$, while that of UGC 12506 is $\lambda=0.15$. Whether we compare to the \HI-derived or the optically-derived sample, the results are the same: UGC 9037 has a typical value, while UGC 12506 has a very high spin parameter. If we compare to the results obtained by \citet{shanHIghMass} for the HIghMass sample as a whole, we find that UGC 9037 has a somewhat low spin parameter, while UGC 12506's spin parameter is still quite high.

\section{Discussion}
\label{sec:discussion}

\subsection{UGC 9037: On the Verge of a Starburst?}
\noindent
The specific star formation rate of UGC 9037 is $\text{SSFR} \equiv \text{SFR}/M_* = 3.5\times10^{-10}$ yr$^{-1}$, which is elevated compared to the typical value for an galaxy with the same stellar mass ($0.9\times10^{-10}$ yr$^{-1}$; \citealt{alfalfa-population}; \citealt{shanHIghMass}). The \HI\ and \Halpha\ both tell the same story in UGC 9037: this is a galaxy which is entering into a phase of significant star formation. The \Halpha\ imaging shows very strong star formation is occurring in the central parts of the galaxy, which is an extremely gas-rich spiral. Across the disk the gas is marginally stable and non-circular flows---possibly inflows---of up to 20 \kms\ are detected. High gas surface densities have built up, fueling star formation: interior of 10 kpc, densities are $>10$ \solarmassespersquareparsec, with the central surface density reaches nearly 14 \solarmassespersquareparsec, significantly above what is typically observed to be the maximum in local star-forming galaxies. All four of our examined star formation thresholds (axisymmetric gravitational instability, shear, formation of a cold phase, and \Htwo\ dominance) point to strong future star formation ability.

But why does UGC 9037 appear to be just now converting this large, dynamically unstable reservoir of gas into stars? The simplest explanation is that the gas is unstable due to its weak bar, but this does not explain why the galaxy has that feature. UGC 9037 is in the outskirts of the NGC 5416 group (poor cluster MKW 12), and so tidal torques from interactions with the group may have created a bar instability in the disk. It is also possible that UGC 9037 has only recently acquired (or is in the process of acquiring) a substantial reservoir of gas, and its dynamics are unable to sustain the gas without an increase in star formation. Both the overall symmetry of the gas distribution and the regularity of the rotation curve indicate that a recent major merger is highly unlikely, but if it is in the process of accreting from an extended gaseous halo or, as it is in a group, cold gas along a filament, then this is what we would expect to observe. However, given the lack of a strong warp in UGC 9037, any accreted gas would have to have the same angular momentum as the parent galaxy (e.g. \citealt{ShenSellwood2006}). Given how difficult it is to determine inflow rates in nearby galaxies, it is unclear as to whether it will be undergoing major star formation and gas depletion in the near future, or whether it is in a quasi-static state of elevated, but not extremely high star formation ($\sim4$ M$_\odot$ yr\per{1}).

\subsection{UGC 12506: High Spin Parameter}
\noindent
In contrast, every property of UGC 12506 points to it as being a massive galaxy with an unusually high spin parameter, which has maintained its gas reservoir in a relatively diffuse configuration, with $\Sigma_\text{HI}$ typically typically $1-4$ \solarmassespersquareparsec. Its gas dynamics are stable, it has a large rotational velocity, and has uniform low-level star formation. While a cold phase of atomic gas can form at modest radii ($\lesssim20$ kpc), our star formation thresholds do not suggest that the \Htwo\ becomes the dominant phase of the ISM except the very center of the galaxy. We observe modest star formation rates across the disk, and similarly only modest levels of gravitational instability in the disk ($\Sigma_\text{HI}/\Sigma_Q\sim0.45$). Our calculation of its spin parameter as $\lambda=0.15$, very high compared to galaxies in the local universe, confirms these observations.


\section{Conclusions}
\label{sec:conclusions}
\noindent
The HIghMass sample is a unique sample of several dozen unusually \HI-rich high mass galaxies out of galaxies detected in ALFALFA with \HI\ masses $>10^{10}$ M$_\odot$. In this paper, we have presented a first look at two of these galaxies using high-resolution VLA observations combined with previous \Halpha\ and optical imaging. We have derived structural and dynamical parameters of the gas distribution and derived dark matter halo properties, including their spin parameters. Future analysis will extend these techniques to the rest of the HIghMass sample.

Our primary question is how galaxies can maintain such large gas reservoirs to the present day without processing them into stars. Of the two galaxies, UGC 9037 appears to be undergoing a period of increasing star formation, possibly stimulated by a bar instability in the inner part of the disk. 
Given the very high ($>10$ \solarmassespersquareparsec) surface densities at the center, we expect to see very high quantities of H$_2$ present there. 
The non-circular flows detected in the disk may stem from radially inflowing gas to fuel the star formation, although flows along the weak bar in this galaxy may be the more likely explanation. Its current global rate of star formation ($\sim4$ M$_\odot$ yr\per{1}) may be sustainable by accretion from its gaseous halo, but if it does undergo rapid star formation in the near future, it may quickly transition toward the red sequence.

The other galaxy, UGC 12506 appears to have a large halo spin parameter ($\lambda=0.15$), leading to lower gas densities and thus star formation rates over the entire disk. UGC 12506's gas is quite stable, but supporting of modest star formation at all radii. We predict its gas to be \HI-dominated at essentially all radii, with relatively little H$_2$. Overall, the most consistent history of UGC 12506 is that its high spin has prohibited it from reaching high rates of star formation, and its baryonic component remains gas-dominated.

\noindent
\textbf{Acknowledgements}\\
\\
This work has been supported by NSF-AST-0606007 and AST-1107390, grants from the Brinson Foundation, and a Student Observing Support award from NRAO.\\
\\
We appreciate the significant help in data reduction of VLA data using CASA provided by our conversations with NRAO staff, especially Emmanuel Momjian and Juergen Ott.\\
\\
We appreciate access to the tilted ring fits and mass models of the galaxies in the THINGS sample as provided to us by the THINGS team.\\
\\
We would like to thank our anonymous reviewer, whose comments have significantly improved this paper.

\begin{centering}
\begin{deluxetable}{lcccccccc}
\tablecolumns{9}
\tablewidth{0pt}
\tabletypesize{\scriptsize}
\tablecaption{Sample Optical, UV, and \HI\ Properties}
\tablehead{
  \colhead{Galaxy} & Optical Coordinates & \colhead{$V_\text{sys}$} & \colhead{dist} & \colhead{$\log\text{M}_\text{HI}$} & \colhead{$\log\text{M}_*$} & \colhead{GF} & \colhead{log SFR}   & \colhead{$R_{25}$}\\
                   & J2000               & \kms                     & Mpc            & M$_\odot$                          & M$_\odot$                  &              & M$_\odot$ yr$^{-1}$ & kpc\\
                   & (1)                 & (2)                      & (3)            & (4)                                & (5)                        & (6)          & (7)                 & (8)
}
\startdata
UGC 9037           & 14:08:29.1+07:03:27 & 5939 & 88.5 & 10.33 & 10.09 & 1.75 & 0.63 & 23.2\\
UGC 12506          & 23:19:30.5+16:04:27 & 7237 & 98.2 & 10.53 & 10.46 & 1.41 & 0.40 & 40.0
\enddata
\tablecomments{\label{tab:sample}\HI\ single-dish, optical, and fitted parameters of the HIghMass galaxies presented in this paper. Column 1: Right ascension and declination of the optical centroid in J2000 coordinates; Column 2: Recessional optical velocity from \citet{alpha.40}, defined as $cz$; Column 3: Hubble flow distance, assuming H$_0=70$ \kms; Column 4: \HI\ mass, from \citet{alpha.40}; Column 5: Stellar mass, from \citet{alfalfa-population}; Column 6: gas fraction, defined as M$_\text{HI}/$M$_*$; Column 7: Star formation rate, from  \citet{alfalfa-population}; Column 8: Optical B-band semi-major axis where surface brightness reaches 25 mag arcsec\per{2}, from \citet{UGC}}
\end{deluxetable}
\end{centering}

\begin{centering}
\begin{deluxetable}{lccccc}
\tablecolumns{6}
\tablewidth{0pt}
\tabletypesize{\scriptsize}
\tablecaption{Map Properties}
\tablehead{
  \colhead{Galaxy} & \colhead{$R_\text{HI}$} & \colhead{$\hat{R}_\text{HI}$} & \colhead{$\log R_\text{HI}/R_{25}$} & \colhead{PA} & \colhead{$i$}\\
                   & kpc                     & kpc                           &                                & $^\circ$     & $^\circ$\\
                   & (1)                     & (2)                           & (3)                            & (4)          & (5)
}
\startdata
UGC 9037           & $42.09 \pm 0.72$ & $43.9 \pm 3.5$ & 0.26 & -11.2 & 60 \\
UGC 12506          & $57.8 \pm 1.9$   & $55.6 \pm 4.5$ & 0.16 & -99.7 & 86
\enddata
\tablecomments{\label{tab:vla-properties}Properties derived from VLA \HI\ maps. Column 1: Radius where $\Sigma_\text{HI}=1$ \solarmassespersquareparsec (see \S\ref{sec:gasmodel}); Column 2: Expected \HI\ radius, calculated from M$_\text{HI}$ using Equation \ref{eqn:Broeils} and \citet{BroeilsRhee1997}; Column 3: Ratio of \HI\ to optical radii (see Table \ref{tab:sample} Column 8); Column 4: Typical position angle on the sky of galaxy in the outer disk (defined as east of north), used to produce position-velocity maps; Column 5: Average inclination angle of the galaxy.}
\end{deluxetable}
\end{centering}


\begin{centering}
\begin{deluxetable}{lcccc}
\tablecolumns{5}
\tablewidth{0pt}
\tabletypesize{\scriptsize}
\tablecaption{VLA Clean Data Cube Properties}
\tablehead{
\colhead{Galaxy} & \colhead{Velocity Resolution} & \colhead{Clean Beam}         & \colhead{Noise (Cube)}       & \colhead{Noise (Map)}\\
                 & \colhead{\kms}          &                              & mJy beam\per{1}              & \\
                 & \colhead{(1)}           & \colhead{(2)}                & \colhead{(3)}                & \colhead{(4)}
}
\startdata
UGC 9037         & 7                       & $7.5\arcsec\times5.7\arcsec$ & 0.52                         &  9.0 mJy \kms\ beam\per{1}\\
                 &                         & $3.2$ kpc $\times 2.4$ kpc   &                              & $2.3\times10^{20}$ cm\per{2}\\
UGC 12506        & 7                       & $8.6\arcsec\times5.8\arcsec$ & 0.46                         & 10.0 mJy \kms\ beam\per{1}\\
                 &                         & $4.1$ kpc $\times 2.8$ kpc   &                              & $2.2\times10^{20}$ cm\per{2}\\
\enddata
\tablecomments{Properties of the VLA data cubes and maps. Column 1: velocity resolution of the image cubes, in \kms; Column 2: the size of the restoring clean-beam, in both arcseconds and kpc; Column 3: Per-channel noise in the cleaned data cubes; Column 4: Noise in the \HI\ intensity maps, both in mJy \kms\ beam\per{1} and in \HI\ column density. As both cubes had identical observational parameters, their clean beam sizes and noise levels are very similar.\label{tab:jvla-res}
} 
\end{deluxetable}
\end{centering}

\begin{centering}
\begin{deluxetable}{llccc}
\tablecolumns{5}
\tablewidth{0pt}
\tabletypesize{\scriptsize}
\tablecaption{\HI\ Spectra-Derived Galaxy Parameters}
\tablehead{
  \colhead{Galaxy} & \colhead{Instrument}  & \colhead{Flux}  & \colhead{W50}     & \colhead{V$_\text{sys}$} \\
                   &                       & Jy km s$^{-1}$  & kms s$^{-1}$      & kms s$^{-1}$ \\
                   &                       & (1)             & (2)               & (3)
}
\startdata
UGC 9037           & VLA                   & $12.68\pm0.34$  & $301.8\pm2.6$     & $5943.0\pm1.8$ \\
                   & ALFALFA               & $11.51\pm0.1$\  & $294\ \ \pm2\ \ $ & $5939\ \ \pm1\ \ $ \\
                   & Difference ($\sigma$) & 3.3             & 2.4               & 1.9\\
\hline\\
UGC 12506          & VLA                   & $14.45\pm0.34$  & $461.6\pm3.0$     & $7235.8\pm1.5$ \\
                   & ALFALFA               & $14.82\pm0.11$  & $457\ \ \pm5\ \ $ & $7237\ \ \pm2.5$ \\
                   & Difference ($\sigma$) & 1.0             & 0.8               & 0.41\\
\enddata
\tablecomments{\label{tab:stats}Comparison of derived quantities from the VLA and ALFALFA spectra of each galaxy (see Figure \ref{fig:spectra}). Derivation of ALFALFA measurements is discussed in \citet{alpha.40}; VLA measurements are derived from spectra in the same way, but with $W50$ measurements corrected for instrumental broadening as discussed \citet{hiarchive}; Column 1: The total integrated flux density of the galaxy; Column 2: The full velocity width at half of the peak emission; Column 3: The systemic heliocentric velocity of the galaxy, using the optical convention. All reported errors are 1-$\sigma$ uncertainties using standard propagation of uncertainty.}
\end{deluxetable}
\end{centering}

\begin{centering}
\begin{deluxetable}{l|ccc|ccc|c}
\tablecolumns{8}
\tablewidth{0pt}
\tabletypesize{\scriptsize}
\tablecaption{Dark Matter Halo Properties}
\tablehead{
  \colhead{Galaxy} &               & \colhead{NFW Halo} &                     &                     & \colhead{Iso Halo} &                     & Spin Parameter\\
                   & $c$           & $R_\text{200}$     & $\chi^2_\text{NFW}$ & $\rho_C$            & $R_C$              & $\chi^2_\text{Iso}$ & $\lambda$ \\
                   &               & kpc                &                     & $10^{-3} \text{M}_\odot/$pc$^{3}$ & kpc         &                     & \\
                   & (1)           & (2)                & (3)                 & (4)                 & (5)                & (6)                 & (7)
}
\startdata
UGC 9037           & $2.67\pm0.50$  & $121\pm12$        & 1.29                & $18.0\pm2.0$        & $5.16\pm0.40$      & 0.65 & 0.07\\
UGC 12506          & $14.87\pm0.60$ & $123.0\pm1.5$     & 0.21                & $1150\pm360$        & $0.91\pm0.15$      & 0.54 & 0.15\\
\enddata
\tablecomments{\label{tab:darkmatter}Results of fitting both Navarro-Frenk-White (NFW) and pseudo-isothermal (Iso) dark matter halo profiles to the two galaxies. Columns 1-2: Concentration index and characteristic length scale of halo (see Equation \ref{eqn:nfw-velocity}); Column 3: $\chi^2$ per degree of freedom of fit to NFW profile; Columns 4-5: central density and core scale length of halo (see Equation \ref{eqn:iso-velocity}); Column 6: $\chi^2$ per degree of freedom of fit to Iso profile; Column 7: Modified spin parameter of the Iso dark matter halo, as discussed in \S\ref{sec:spin-derive}.}
\end{deluxetable}
\end{centering}

\end{document}